**Title**

# The potential for solar-diesel hybrid mini-grids in refugee camps: A case study of Nyabiheke camp, Rwanda

**Author names and affiliations**

Javier Baranda Alonso[1], Philip Sandwell[1,2,3], Jenny Nelson[1,2]

[1] *Department of Physics, Imperial College London, London SW7 2AZ, UK*

[2] *Grantham Institute - Climate Change and the Environment, Imperial College London, SW7 2AZ, UK*

[3] *Practical Action, Rugby, CV21 2SD, UK*

*Author emails: javier.baranda-alonso18@imperial.ac.uk, philip.sandwell09@imperial.ac.uk, jenny.nelson@imperial.ac.uk*

**Corresponding author**

Javier Baranda Alonso

*Email: javierbarandalonso@gmail.com, javier.baranda-alonso18@imperial.ac.uk*

*Postal address: Department of Physics, Imperial College London, London SW7 2AZ, UK*

**Abstract**

Electricity access in refugee camps is often limited to critical operations for humanitarian agencies and typically powered by fossil fuel generators. We study the economic and environmental benefits that optimised fully renewable and diesel-hybrid mini-grid designs can provide in humanitarian settings by displacing diesel use. Considering the case study of Nyabiheke camp in Rwanda we found that these benefits are substantial, with savings up to 32% of total costs and 83% of emissions, and cost payback times ranging from 0.9 to 6.2 years. Despite of their different cost structures, we find that all hybridisation levels of the system provide cost and emission savings compared to the incumbent diesel system. We highlight how modelling tools can facilitate the introduction and progressive expansion of systems as well as inform operational considerations on the ground. This study demonstrates how financial resources, environmental objectives and operational timeframes will influence the most appropriate system design for humanitarian actors on a case-by-case basis.

**Highlights**

- Renewable mini-grids can provide substantial benefits in humanitarian settings
- Solar-diesel hybrid mini-grids are a cost-efficient solution to displace diesel use
- Optimal hybridisation level depends on available resources for humanitarian agencies
- Sustainable mini-grids can save up to 32% of total costs and 83% of total emissions
- Payback times ranging from 1 to 6 years can fit short-term organisational budgets

**List of Abbreviations**

| | |
|---|---|
| CLOVER | Continuous Lifetime Optimisation of Variable Electricity Resources |
| CRRF | Comprehensive Refugee Response Framework |
| GHG | Greenhouse gas |
| MAC | Marginal Abatement Cost |
| LCOE | Levelised Cost Of Electricity |
| LCUE | Levelised Cost of Used Electricity |
| PPA | Power Purchase Agreement |
| ROI | Return On Investment |
| RWF | Rwandan Franc |
| UNHCR | United Nations High Commissioner for Refugees |
| UNITAR | United Nations Institute for Training and Research |

## 1. Introduction

### 1.1. Humanitarian energy: Energy provision in humanitarian contexts

In recent years the world has experienced continuously rising humanitarian needs and record numbers of displaced people, reaching 70.8 million globally in 2019. Of those, 25.9 million refugees have sought asylum in outside of their home countries, escaping from conflict, prosecution and climate-related events. Developing countries continue to disproportionally bear the largest responsibility of hosting refugees, with 84% being hosted in developing regions [1,2]. The funding shortage to address increasing humanitarian needs [1], together with lack of resources on many refugee-hosting countries, further hamper the identification of durable and sustainable solutions for displaced people, especially in protracted situations [3].

Access to energy plays a crucial role in ensuring the provision of basic services and essential economic activities in developing countries [4–6], as well as in humanitarian settings [3]. The main energy uses present in displacement settings can be categorised in public use (institutional and operational energy requirements), household use (provision of energy for household lighting, heating and cooking) and productive uses (energy requirements for commercial and income-generating activities) [7]. Institutional loads refer to the energy requirements to power humanitarian agencies and offices in displacement settings. Operational energy needs comprise the provision of basic services through water pumping, health centres, educational centres or street lighting, generally managed by humanitarian actors [8].

Increased levels of energy access in humanitarian settings bring clear benefits both for displaced populations and humanitarian organisations. These are reflected in increased levels of security, better access to health services, improved livelihoods, opportunities and resilience, fostering local economic activity and bringing local environmental benefits [3,9,10].

However, energy provision in displacement settings is typically limited to the minimum requirements for survival and critical operations of humanitarian actors: it is estimated that almost 90% of refugees in displaced settings globally do not have access to electricity [3]. Furthermore, energy has not been recognised as a priority in humanitarian assistance, resulting in a historically poor working knowledge and a fragmented approach towards energy supply interventions in the humanitarian sector [3,10,11]. In parallel, the regulatory uncertainty around displaced settings has prevented displaced populations from being successfully included in national energy-access agendas. The energy needs of refugees remain poorly understood and inadequately integrated into energy supply interventions in many cases [12].

Only recently has energy access emerged as a relevant factor in the provision of humanitarian assistance, associated with an increasing environmental awareness in the sector and a willingness to better understand its energy usage and reduce its environmental footprint [13,14]. Different international initiatives aim to address energy needs in displacement and humanitarian settings specifically, both providing local solutions as well as setting a common operational framework for the sector to transition to cleaner energy solutions. Among them are the Moving Energy Initiative [9], the UNITAR-led Global Plan of Action for Sustainable Energy Solutions in Situations of Displacement [10], or the UNHCR Global Strategy for Sustainable Energy [13].

Rwanda, a densely populated landlocked country in East Africa, presents a forward-looking example towards rural electrification and the integration of displaced populations. Despite its relatively small size, the country currently hosts 150,000 refugees, primarily from the Democratic Republic of the Congo (DRC) and Burundi [15]. Rwanda is an adopter of the major refugee conventions worldwide, including the New York Declaration for Refugees [16] and its Comprehensive Refugee Response Framework (CRRF) [17]. These conventions aim to provide long-term sustainable solutions for refugees and to integrate them in the socio-economic life of the host countries. The country upholds a progressive regulatory framework towards the status of refugees, allowing refugees to work and access other national services such as healthcare and education [18,19].

With regards to its rural electrification efforts, Rwanda's ambitious electrification strategies [20,21], favourable policies towards off-grid electrification and effective financing programmes have resulted in a rapid scale-up of electricity access in the country, rising from 8% of households in 2008 to 35% in 2017 [22,23]. Despite the favourable environment and the presence of specific initiatives targeting energy access in displacement settings, such as the Renewable Energy for Refugees project [24], energy access in refugee camps in Rwanda is still very limited [25]. There remains a need for continued investment and efforts to sustainably meet the energy needs of displaced population in the country [26,27]. Nevertheless, the progressive national framework and a vibrant solar off-grid private sector offer an exceptional opportunity to implement sustainable energy solutions in the existing refugee camps in the country [28–30].

### 1.2. Challenges for cleaner energy supply

The increasing humanitarian crisis and chronic funding shortages have limited the capacity of humanitarian organisations to deploy modern and efficient solutions for energy supply in displacement settings. Humanitarian actors have traditionally relied on inefficient, polluting and expensive stand-alone diesel generators for electricity supply in remote displacement settings where the grid supply is not available [8,11]. This practice is due to the widespread availability and maturity of the technology, in addition to the ability to provide a continuous and reliable electricity supply, and to be rapidly deployed in most locations with relatively low capital expenditure. However, the diesel systems installed are often oversized for the loads required, resulting into lower load factors and poor fuel efficiencies. Together with the high associated operational costs and the exposure to fuel price volatility, this poses several economic and logistical challenges to humanitarian organisations, in addition to the health impact caused by local air pollution in camps. The Moving Energy Initiative estimates that humanitarian agencies spend up to 5% of their annual budget on fossil fuels, including electricity generation in camps and for transportation, amounting to $1.2 billion annually across the sector [11].

Considering this economic and environmental burden it is therefore imperative to transition to cleaner and more cost-effective energy sources in humanitarian interventions. This aligns with the "no harm" principle of humanitarian work in an effort to minimize the environmental footprint of humanitarian assistance [10,11,13].

Humanitarian organisations, donors, host governments and other actors are increasingly pledging to use renewable energy to improve the situation [13,14]. Nonetheless, different barriers have hampered the introduction of clean energy infrastructure in displacement settings [3,10,31]:

- Lack of in-house technical expertise, scarce data on energy use both of displaced people and institutions, and a lack of comprehensive and coordinated strategies for energy provision and management of humanitarian organisations.
- High upfront costs of renewable energy infrastructure, which poorly matches limited and short-term funding cycles of humanitarian actors, which generally operate on an annual basis.
- The perception that protracted situations are short-term issues, regulatory uncertainty around the status of displacement settings, the perceived risk of investing in long-term energy infrastructure, and the marginalisation of lack of acknowledgement of displaced populations in national electrification agendas.

In order to successfully deploy and scale-up sustainable energy solutions in refugee camps, continuous support in the form of favourable policy frameworks and increasing engagement of the private sector are crucial to facilitate the financing and long-term management of renewable assets [8,31]. However, organisations willing to invest in clean energy infrastructure still face high risks and lack the technical and financial resources to transition towards renewable energy sources completely.

### 1.3. Opportunities for mini-grid systems

Decentralized off-grid clean energy systems, such as solar PV and battery storage mini-grids, have emerged as sustainable and reliable solutions for energy provision in rural areas where the grid extension is not physically or economically viable [22,32,33]. Given the constraints present in humanitarian settings, and to better adapt to the objectives and resources available for humanitarian organisations, a wide range of solar-diesel hybridised mini-grid systems can provide a significant reduction in operational costs and environmental impact with reduced additional investments, taking advantage of the existing diesel infrastructure [34,35]. Hybrid solar-battery-diesel mini-grid systems, onwards referred as 'hybrid systems' in this study, benefit from the flexibility and reliability of traditional diesel generators as well as the reduced requirements of renewable generation and storage capacity installed in comparison with fully renewable systems [36]. Thus, hybrid systems can represent a more cost-effective solution than fully renewable systems in humanitarian contexts.

As well as providing critical services, humanitarian assistance also aims to improve the livelihoods of displaced populations and host communities and catalyse economic growth pursuing long-term sustainable and resilient development. The introduction of renewable energy infrastructure offers the possibility of expanding the electricity supply beyond the basic necessities. This could provide a reliable, safe and affordable electricity supply to entrepreneurs and local businesses for the development of productive activities [37].

The unique context of displacement settings offers a favourable environment for the electrification of productive activities through mini-grid systems. The high population density and concentration of businesses represent a large base of potential customers. Furthermore, reliable high-consuming anchor loads can increase the profitability of mini-grids, with the institutional and operational loads present in displacement settings or refugee camps operated by trustworthy and bankable organisations, potentially providing revenue certainty for private mini-grid suppliers.

However, due to the short-term nature of the humanitarian response and the lack of funding and comprehensive long-term strategies in many interventions, energy systems deployed in these settings are often planned over relatively short timeframes and designed to meet a certain level of electricity demand. In parallel, the lack of financial and technical capacity from humanitarian agencies mean that these systems might be in operation for significantly longer periods than the ones initially intended, often without adequate maintenance or replacement of the degraded capacity. They might also experience unexpected growth in electricity demand over time, for example as a result of the connection of additional institutional loads, operational loads, households or productive users. It is therefore important to consider these factors when planning the design of electricity systems, and their implications on system performance, to ensure their suitability over their operational lifetime.

Finally, while providing a cheaper and cleaner electricity supply for humanitarian organisations and refugees alike is a crucial step towards the promotion of equitable livelihood opportunities for displaced populations, these interventions must be financially accessible and sustainable. Thus, the establishment of suitable delivery models and contractual mechanisms to engage the private sector in the implementation and maintenance of such systems is a key enabler of the successful scale-up of sustainable energy solutions in humanitarian settings. These mechanisms have the potential to facilitate facing the investment required and mitigate the risks associated with the operation and maintenance of the systems [8,38].

### 1.4. Modelling mini-grid systems in humanitarian situations

To maximize the benefits that decentralised renewable systems can offer, the accurate sizing and modelling of the system remains a crucial factor of success. This applies both during initial planning phases as well as when considering the long-term performance and potential expansion of systems. This modelling process frequently aims to determine the minimum system requirements needed to meet the existing electricity demand under specific criteria, usually associated with the system cost or the reliability of the electricity supply [39,40]. Different modelling and optimisation tools and techniques have been used in similar rural off-grid contexts, with the use of software packages for the design and analysis of power systems becoming a widespread solution [39,41–43].

Political and international support to design and introduce more sustainable solutions for displaced populations, including the expansion of the energy provision through cleaner energy sources has extended recently [10,13,17,24], but critical analysis of the existing options for practical implementation has not been widely explored:

- The quantification of the economic and environmental costs of prevalent assumptions about hybridising mini-grids and their benefits is required to compare them objectively with incumbent diesel-powered and fully renewable systems.
- Modelling and optimisation tools to investigate different strategies and system design approaches can help to inform solutions tailored to the needs and resources of humanitarian organisations in each specific context.
- Successful and detailed case studies are needed to build an evidence base around the potential of sustainable mini-grid systems in humanitarian settings, supporting and informing strategies for large scale rollout of hybridisation in the humanitarian sector.

Addressing these three issues will contribute towards a greater understanding of how to provide sustainable electricity in situations of displacement.

### 1.5. Summary of the study

This study explores the potential strategies of introducing renewable technologies (solar PV and battery storage) to partially or fully displace diesel use for humanitarian operations in refugee camps and the benefits that they entail, considering the current context and constraints faced by the humanitarian sector.

The institutional and operational electricity demand existent in humanitarian settings is presented through a representative case study, here being Nyabiheke refugee camp in Rwanda. This demand comprises of institutional loads, operational loads and productive loads corresponding to refugee businesses. To represent the expansion of the electricity supply to productive users in the camps, we explore different productive load expansion scenarios.

Using the CLOVER optimisation tool [44,45], we assess the impact of the use of incumbent diesel generators to meet the existing electricity supply requirements in Nyabiheke. Different sustainable mini-grid system designs are optimised in CLOVER for the current energy demand in Nyabiheke. These correspond to the lowest levelised cost of used electricity (LCUE) over its lifetime for a fully renewable PV-battery and multiple hybrid PV-battery-diesel systems, presented together with their economic and environmental implications. Additionally, we present the optimal system design for different levels of diesel hybridisation, corresponding to different renewable fractions and identifying short and long-term economic and environmental impacts.

We identify and discuss the correspondence of different hybridisation levels with the objectives and resources available for humanitarian organisations and the advantages that such systems can offer in displacement contexts. We provide system design recommendations and analyse approaches for extending the energy supply to additional productive users in refugee camps. Finally, we present different alternatives for the deployment of the previous sustainable mini-grid systems in cooperation with the private sector.

We conclude by highlighting the main implications on existing policy frameworks of host countries and current deployment strategies of humanitarian organisations. Drawing from the previous analysis, we present suggestions and best practices for policy-makers, private actors and humanitarian organisations to maximise the impact and scalability of sustainable mini-grids in displacement settings.

## 2. Methodology

### 2.1. Description of modelling tool: CLOVER

Given information about the electricity demand and the energy resources available for a closed system, in this case Nyabiheke camp, we use the optimisation software CLOVER for the design and analysis of optimised mini-grid systems according to specified economic and operational criteria. CLOVER (Continuous Lifetime Optimisation of Variable Electricity Resources) is an open-source energy systems simulation and optimisation tool designed for supporting rural electrification in developing countries. Similarly to other widely commercialised software as HOMER [42,46], CLOVER allows performing hourly simulations of selected energy systems, providing indicative performance metrics, as well as optimising system component sizes to meet predetermined objectives. A comprehensive description of the CLOVER model and capabilities can be found in [44,47].

The optimisation process, characteristic of CLOVER, involves both optimisation and sufficiency criteria. CLOVER performs series of simulations with increasing system sizes until the performance of one of the systems simulated meets the sufficiency criterion defined by the user, for instance, a minimum level of reliability in the electricity supply. CLOVER then analyses surrounding systems in terms of component sizes, looking for systems that both meet the sufficiency criteria and perform better according to the optimisation criteria, for instance at a lower cost. This process is repeated until adjacent systems are no longer superior, selecting the last system as the optimum and recording its environmental, technical and economic performance. Having performed this for a given time period CLOVER then repeats the process for the next time period, and so on for the entire system lifetime, allowing the optimum system size to increase in response to technological degradation or growing load profiles whilst maintaining the original performance. This allows for long-term planning whilst replicating the responsive strategies for system upgrades that likely would be implemented in practice in reaction to changing performance.

Regarding the electricity load profiles used for the optimisation and simulation processes, CLOVER allows to define them in two different ways. The model allows to create a load profile based on information about appliance availability and demand for their services, provided for instance through survey information. Alternatively, a representative load profile, obtained for example from monitored usage data, can be directly input, as done in this study.

### 2.2. Data collection and electricity demand scenarios

Most of the analysis of this study relies on primary electricity load data collected in Nyabiheke Refugee Camp, Rwanda, between March and June 2019. These data were collected by Practical Action [24] through smart meters monitoring existing loads connected to the incumbent diesel-powered system in place, including two water pumping stations, a health centre, various blocks of institutional offices and several privately-owned businesses by refugees. This electricity load profile comprises the Baseline scenario analysed in this study. In order to evaluate the economic, environmental and technical performance of the incumbent and the proposed systems, the required modelling parameters, including financial and environmental inputs for the study were sourced from relevant literature.

Three additional electricity demand scenarios are also considered in later sections of this study to evaluate the implications of expanding the electricity supply in Nyabiheke further than the current Baseline scenario to productive users and refugee businesses within the camp and immediate surroundings. This consideration is in line with the existing efforts of governments and organisations to promote the socio-economic activity and integration in displacement settings through electricity provision. We consider three expanded productive demand scenarios representing different levels of electricity provision for productive activities. In the *Low Productive scenario*, the final productive electricity load equal to three times the current productive electricity load in the Baseline scenario. In the *Medium Productive scenario*, the final productive electricity load equal to six times the current productive electricity load in the Baseline scenario. Finally, in the *High Productive scenario*, the final productive electricity load equal to ten times the current productive electricity load in the Baseline scenario.

We illustrate the practical implications of the different levels of expansion of the electricity supply through the number of businesses that could be powered in each of the scenarios. As the requirements and usage of individual businesses will vary in reality, the total number of businesses is offered as an indication on the potential scale and scope of the overall demand and its possible uses, rather than an exact estimation of the number of connections, for example. The electricity demand for each business is estimated following the World Bank Tier 2 and Tier 3 levels of electricity demand [48], according to the objectives presented in the UNHCR Strategy for Sustainable Energy [13].

### 2.2. Analysis criteria and metrics used

Various metrics are utilised in this study to characterise and evaluate the technical, economic and environmental performance of the mini-grids systems being evaluated. These are described in this section and the equations used to calculate them are available in the Supplementary Information.

The **system reliability** is defined as the proportion of the number of hours during which power is available compared to the total number of hours of operation of the system. A loss of power, referred to as a blackout, occurs when the energy demanded from the load exceeds the capability of the system to meet it, resulting in a temporal shortage of energy. The system reliability is a crucial metric to evaluate the quality and security of the electricity supply, especially where there is presence of critical loads, such as health facilities

The **Levelised Cost of Used Electricity (LCUE)** is defined as the discounted cost per kWh of electricity used to meet the demand over the lifetime of the project. In comparison with the traditional levelised cost of electricity (LCOE), the LCUE explicitly does not account for the electricity generation that cannot be used due to reduced demand and limited electricity storage capacity, being dumped, and accounts for shortfalls in supply and reliabilities less than 100%. Thus, the LCUE represents more accurately the cost of electricity assumed in off-grid systems to meet demand, and is used as the main economic indicator in this study.

The **renewable fraction** of the system is defined as the proportion of the total energy provided by renewable sources over a certain period or lifetime compared to the total energy supplied by the system. In our case this can range from 0 (a diesel-only system) to 1 (a fully renewable solar and storage system) or any value in between (a hybrid system). This will be used as a key metric to identify systems with different levels of hybridisation of the energy supply.

The **cumulative or embedded greenhouse gas (GHG) emissions** refers to the total GHG emissions associated with the production and operation of the system over its lifetime. This metric is used to assess the environmental impact of different systems, and calculated in terms of kgCO2eq [49].

The **emissions intensity** of an electricity source refers to the cumulative GHG emissions released per kWh of electricity supplied by the system, allowing to compare the specific GHG emissions of different energy systems between them.

The **Marginal Abatement Cost (MAC)** refers to the financial costs associated with the mitigation or reduction of a negative impact, in this case, the GHG emissions associated with the baseline diesel system. For instance, a negative MAC implies that the alternative studied presents both lower costs and emissions associated. Therefore, the MAC is used in this study as a metric to compare the cost-effectiveness of mitigation emissions by different systems.

As mentioned in Section 2.1, CLOVER selects the optimal energy system characteristics based on the sufficiency and optimisation criteria. The sufficiency criteria used for this study is a minimum level of reliability of the power supply by the system over its entire lifetime, in this study 95%, for both hybrid and fully renewable systems analysed. While higher reliability levels would be desirable, achieving increasingly higher levels of reliability for PV-battery-based off-grid systems requires disproportionally larger system capacities, consequently reducing the cost-efficiency of the system and its investment attractiveness [50]. The main optimisation criterion used in this study is the LCUE of the system over its lifetime, due to the possibility to compare systems of different nature as well as setting the electricity tariff required to break even at the end of the lifetime of the system for potential private actors involved in the deployment of the system.

Other criteria used to select the most appropriate systems are the initial or upfront costs required to deploy the system, together with the renewable fraction of the system and the emissions intensity associated, depending on the specific resources and objectives of humanitarian organisations in this regard. For comparability of results, all the costs presented in this study are shown in 2020 US dollar equivalents ($).

### 2.3. Analysis of sustainable mini-grid systems

To mitigate and replace the use of diesel in off-grid systems in humanitarian settings, we consider solar PV as the alternative generation source, supported by energy storage through lead-acid batteries. Given the presence of already deployed diesel generators in most humanitarians settings, we evaluate the potential that both fully renewable and diesel-hybrid mini-grid systems present to reduce the economic and environmental impact of the electricity supply in camps. We use CLOVER to model the performance of multiple mini-grid designs for the case of Nyabiheke, comparing them to the performance of the incumbent diesel system in place.

To this end we simulate and characterise the incumbent diesel system, which serves as a baseline for the analysis of the subsequent designs. We then use CLOVER to find the optimum fully renewable and hybrid systems under selected criteria presented at the end of Section 2.2. In order to compare the performance of the incumbent diesel-powered system with the alternative sustainable mini-grid designs presented in this study, we established a common study framework. This framework involves the main simulation parameters used for the analysis in CLOVER, such as the system lifetime modelled, set as 15 years, or the re-assessment modelling period, considered as 5 years. Despite the temporary nature of displaced settings, we selected these timeframes considering the relatively long periods that refugees can live in these settings [51], particularly in protracted refugee crisis such as the one existing in Rwanda. The 5-year re-assessment period is selected considering the frequently limited capacity of humanitarian organisations to re-assess the system and replace components after its deployment. The remaining simulation parameters are available in Table A.1 in the Appendix. Similarly, the technical, financial and environmental parameters used in this study are included in table A.2. Under this framework, we simulate the incumbent diesel system in Nyabiheke over a lifetime of 15 years with a reliability level of 95%, and compare it with the optimised fully renewable and hybrid systems. The load corresponds to the existing load in Nyabiheke or Baseline scenario, and for simplicity it is considered constant over time without seasonal nor yearly variations.

However, these optimised system designs represent only two of the wide range of hybridisation levels possible, some of which could potentially fit better with the economic resources and environmental objectives of humanitarian organisations in place. We therefore present a sensitivity analysis of the renewable fraction of the system, comparing the optimum system sizes for different levels of hybridisation. This spectrum of systems could allow organisations to identify the system design that aligns best with their interests, considering factors such as the capital investment required, the LCUE of the system, or the GHG emissions offset.

In addition to the economic and environmental performance, the level of hybridisation of the system also brings further operational challenges for the operation of diesel generators in place.

These factors include the number of hours of operation needed and the seasonal variability of the diesel supply required as a consequence of the variable renewable resource throughout the year. These factors are also considered in the sensitivity analysis, highlighting the potential of modelling techniques to inform operational considerations on the ground.

From the range of systems analysed for the specific case of Nyabiheke, we suggest three possible hybridisation levels to match different levels of financial resources available to humanitarian organisations.

### 2.4. Implementation of sustainable mini-grids in humanitarian settings

Despite of the opportunities that the introduction of renewable infrastructure can create in humanitarian settings, the overstretched resources of humanitarian organisations present several challenges for the design and sustainable operation of renewable systems. The short-term vision of humanitarian response can lead to a lack of capability to maintain and expand renewable systems appropriately over longer timeframes and considering future demand growth.

We use CLOVER to evaluate the economic and operational implications of this short-term design approach. For this purpose, we simulate the connection of an unexpected growing load over 15 years to different renewable fraction systems initially optimised to supply a constant load over 5 years (corresponding to the baseline scenario), without any re-assessment or re-sizing of the system during its lifetime. We then compare this short-term approach with two different long-term approaches that account for the productive load growth connected over the system lifetime. A first static approach considers that the system is deployed in its entirety in a one-off installation. A second modular approach involves a system that is re-assessed and re-sized every certain period to adapt to the growing demand, introducing the required generation and storage capacity to meet the expected load growth during each period.

As introduced in Section 1.3, the engagement of the private sector in the implementation of renewable infrastructure in humanitarian settings is key to support the resource-constrained agencies and to scale-up sustainable energy solutions in these contexts. To complement the previous techno-economic analysis presented in this study, a discounted cash flow analysis for different contractual mechanisms to introduce optimised mini-grid systems for the Baseline scenario in Nyabiheke. This cash flow analysis evaluates the costs assumed by humanitarian organisations, highlighting the viability of these contractual mechanisms to deploy renewable infrastructure to displace diesel in humanitarian settings.

We evaluate three of the main contractual mechanisms available for humanitarian organisations:

- *Purchase by a humanitarian organisation*, where the organisation owns and operates the renewable assets, in addition to operating the diesel generator and sourcing the diesel fuel.
- *Lease to own*, where the organisation leases the system from an energy services company, which is responsible for installing, operating and maintaining the renewable system during the lease in exchange of a monthly fee. The organisation is still in charge of operating and sourcing fuel for the diesel generator system, if one is used. The ownership of the system is transferred to the organisation after the end of the lease.
- *Limited Power Purchase Agreement (PPA)*, where the organisation agrees with a solar services company the outsourcing of the energy supply except the operation of the diesel generators. The energy services company installs, operates and maintains the system during the PPA agreement, after which the organisation can renew or cancel the agreement. If cancelled, the renewable assets are removed from the facility.

More details about the specifics of each mechanism and comparisons between them are available in Supplementary Information.

## 2.5. Limitations of the study

This research aims to provide an informative overview of the possibilities of energy systems modelling techniques to design sustainable mini-grid systems to improve the electricity provision in humanitarian settings, using the case study of Nyabiheke camp in Rwanda. The results we present are conditional on the limitations of the data and the analysis tools used, and more in-depth assessments would be needed to inform the deployment of a specific system in Nyabiheke camp or similar displacement contexts. Among the limitations that arise from the analysis, we can highlight three: load data used, technology choice and parameters, and limitations of CLOVER model

First, the load data used for the study is valuable primary data not widely available in humanitarian contexts but was collected for a relatively short period of time and thus lack the duration needed to represent events such as demand growth over time or demand seasonality.

Secondly, the modelling inputs and parameters selected for the analysis have a direct impact on the results and therefore, on the most suitable mini-grid system design. These inputs include the technology choice and cost parameters. An example is the battery technology selected. For this study we consider lead-acid batteries, but using alternative technologies would impact the optimum system design, total storage requirements, and impacts such as the lifetime costs. For instance, the use of lithium ion batteries could present a trade-off between the higher upfront costs per capacity and the improved performance and lifetime, that could allow reducing the storage capacity needed over the system lifetime. Another example is the influence of different diesel fuel costs, considering that displacement settings are often located in remote areas where the regular supply of fuel might be challenging and expensive, with costs significantly over market prices. In this study, these factors would likely favour the introduction of larger fractions of renewable generation in the optimum system designs by making diesel supply more expensive and, as a result, renewable systems more cost-competitive. Therefore, to demonstrate the high potential of sustainable mini-grid systems to displace diesel generation, this study addresses a more conservative scenario, with constant and relatively affordable diesel fuel prices. Moreover, the high electricity demand over night in Nyabiheke, due to the water-pumping requirements, also presents a more favourable scenario for diesel in comparison with PV and battery storage.

Finally, the scope of the study is also constrained by the limitations of the tools used for the analysis. For instance, the analysis in CLOVER does not include the geospatial distribution of the loads in Nyabiheke. While less relevant for the institutional loads, which are relatively concentrated in the camp and already connected to the incumbent diesel-powered mini-grid system, it can represent a major factor towards the profitability of connected more dispersed productive users around the camp. Additionally, the approach used in CLOVER to simulate diesel usage operates diesel generators to fill any unmet demand left by renewable generation and battery storage upon the desired reliability requirements, and thus might not accurately mimic the operation of some real diesel backup systems. Similarly to other modelling tools, the demand in CLOVER is assumed to be known over the entire system lifetime for optimisation and simulation purposes, which contrasts with the high uncertainty and lack of demand data in humanitarian contexts.

## 3. Results and Discussion

### 3.1. Electricity demand in and incumbent diesel system in Nyabiheke refugee camp

As introduced in Section 1, the primary electricity demand present in displacement settings comprises the institutional and operational loads managed by humanitarian organisations, potentially also including other productive activities and businesses owned by refugees and, occasionally, refugee household loads. Figure 1.a shows the existing loads connected to the incumbent diesel-powered system in Nyabiheke refugee camp, Rwanda.

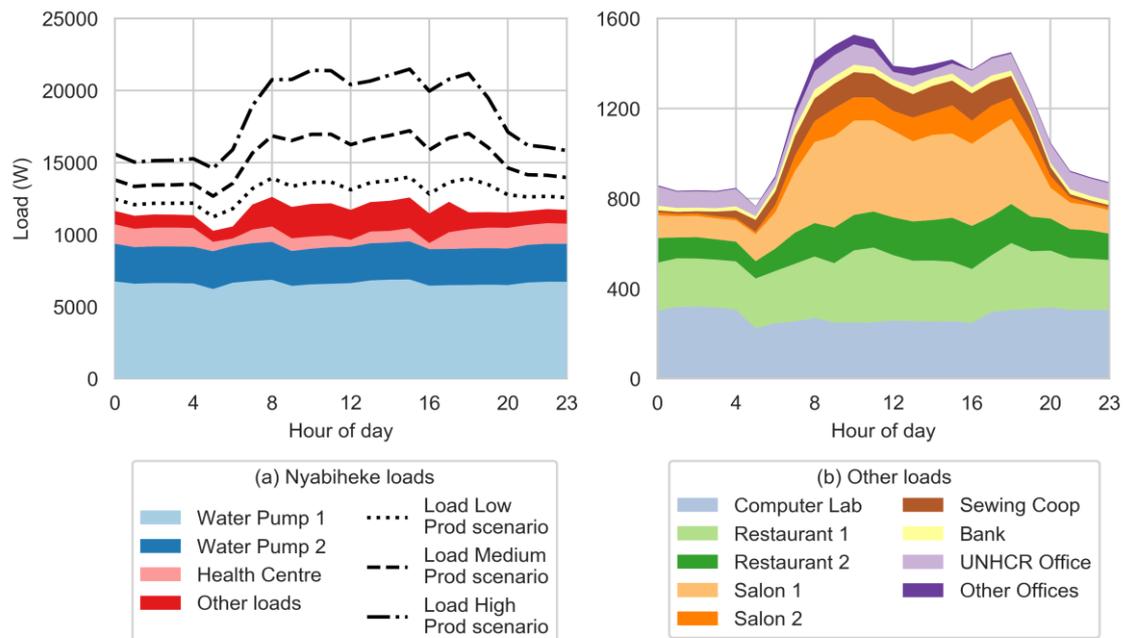

*Figure 1. (Double column color artwork)* Hourly average load connected to the incumbent diesel system in Nyabiheke. Part (a) represents the cumulative load connected to the system (filled colours) and the total load when a given multiple of the current productive load profile, coloured red and detailed in part (b), is added on to the current total (dashed lines).

The institutional and operational loads connected include two water pumping stations, a health centre, an office block and various small administrative buildings. The total electricity demand is dominated by the two water pumping stations in place, accounting on average for more than 9 kW of the approximately 11 kW of the total load connected to the system, due to the continuous pumping requirements to meet the humanitarian emergency water supply standards [52]. These are followed by the health centre, with loads varying from 500 W during day hours to 1.5 kW at night-time. Figure 1.b shows the smaller institutional loads, including office blocks and a bank. It also illustrates the existing refugee productive users connected to the system, which are currently being provided with free electricity at the expense of humanitarian organisations. These include two hair salons, two restaurants with adjacent shops, a sewing cooperative and a computer lab. The total productive demand is concentrated during daylight hours but does not exceed 1.4 kW, thus significantly lower than the institutional electricity demand.

As a result of the continuous water pumping requirements, the overall electricity demand presents a reasonably constant profile over the entire day. This type of demand profile is well suited for diesel generators, which perform best under continuous load requirements and high load factors. On the contrary, the mismatch between this constant load profile and the solar generation profile will translate into higher storage requirements to effectively meet the night-time demand through an entirely renewable system. This characteristic load profile suggests that the introduction of a PV-battery-diesel hybrid system in Nyabiheke, thus reducing these storage requirements, could be a viable solution.

Figure 1.a also shows the total electricity demand for the three additional scenarios presented in Section 2.2. Recent initiatives by humanitarian organisations aim to expand the electricity provision to displaced populations, with a particular focus on promoting the connection and development of productive activities. Table 1 represents the number of businesses that could be powered in each proposed scenario. From section 3.2 to section 3.6, and in section 3.7, we analyse different mini-grid designs for the Baseline scenario. In section 3.6 we focus on the High Productive scenario, identified as the aspirational level of business electrification in Nyabiheke.

| Scenario | Objective | Equivalent number of business supplied Tier 2 (200 Wh/day) | Equivalent number of business supplied Tier 3 (1000 Wh/day) |
|---|---|---|---|
| *Low Prod* | Initial level of business electrification | 185 | 37 |
| *Medium Prod* | Intermediate level of business electrification | 460 | 92 |
| *High Prod* | High level of business electrification | 830 | 166 |

*Table 1.* Equivalent number of additional Tier 2 or Tier 3 businesses that could be powered in each proposed productive load scenario.

Higher number of businesses connected results in progressively more pronounced electricity demand profiles during day-time hours, which concentrate most economic activity in camps. The number of new businesses connected considered in each scenario is considered reasonable given the population of Nyabiheke refugee camp, hosting more than 13,000 refugees, and the vibrant organic economic activity present despite the current lack of energy access, which could significantly increase business opportunities.

Following the study framework presented in Section 2.3, we simulate the performance of the incumbent diesel system in Nyabiheke under the Baseline scenario load over a lifetime of 15 years considering a 95% reliability level. A minimum of 13 kW of diesel generation capacity is required to meet the load, and its economic and environmental characteristics are summarised in Table 2. As expected, the cost structure of the system over its lifetime is heavily dominated by operating diesel fuel costs, accounting for 85% of the total lifetime discounted costs and responsible for 98% of total lifetime GHG emissions, that amount to 103.6 tCO2eq annually.

### 3.2. Optimisation and performance of sustainable mini-grid systems

A fully renewable PV-battery system and a hybrid PV-battery-diesel system are optimised in CLOVER to meet the Baseline scenario demand while minimising the LCUE of the system over its lifetime with a reliability level of 95%. Although the fully renewable system would be the environmentally most sustainable solution, a hybrid system is also considered as humanitarian organisations may require a back-up generation source to ensure the reliability of supply to critical loads such as health facilities. We optimise the systems in CLOVER without accounting for the existence of the 13 kW diesel generator presented in Section 3.1, providing a level playing field for comparisons between systems. The main characteristics of the three types of system considered, i.e. the incumbent diesel based system, the fully renewable system and the hybrid system, are listed in Table 2. Figure 2 represents the hourly energy performance of the cost-optimised fully renewable and hybrid systems, identifying the average hourly fraction of demand met by diesel, solar PV and battery storage.

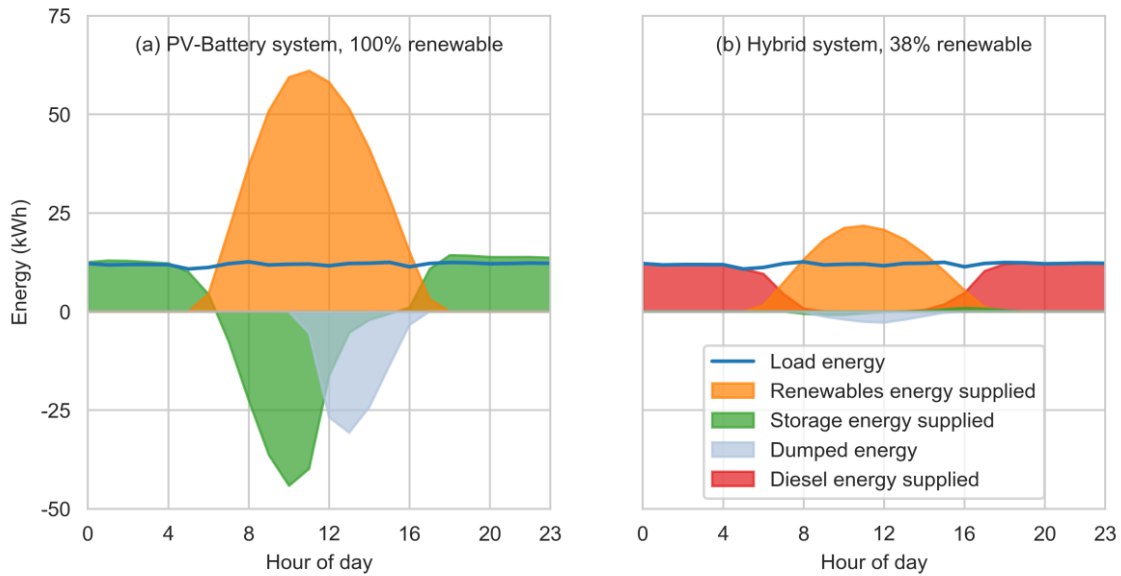

*Figure 2. (Double column color artwork) Average hourly energy performance of optimised mini-grid systems for Baseline scenario load in Nyabiheke refugee camp over their lifetime. (a) Optimised fully renewable PV-battery system with a renewables fraction of 1. (b) Optimised PV-battery-diesel hybrid system with a renewables fraction of 0.38.*

For the fully renewable system, with a renewable fraction of 1, 100 kWp of PV and 410 kWh of storage capacity are required to meet the demand over its lifetime. Figure 2.a shows that solar generation meets the electricity demand during day whilst charging the batteries that supply the demand during night periods. PV and storage capacity at the beginning of each re-assessment period are oversized to account for the degradation of the system over time. Figure 2.a illustrates this with large amounts of dumped electricity, that cannot be stored in the batteries after they are fully charged and referred to as overgeneration, present in the afternoon. The cost structure of the system is dominated by the upfront investment in PV and battery storage capacity, with new equipment costs accounting for 83% of total lifetime costs of the system. Similarly, and due to the complete displacement of diesel use, 84% of total lifetime GHG emissions are associated with the embedded emissions from the manufacture of the installed equipment.

For the hybrid system, the resultant cost-optimised renewable fraction is 0.38 is achieved through 30 kWp of solar PV generation with no initial battery storage capacity installed. Only 10 kWh are installed after the first re-assessment period and due to the reduction in battery storage costs over time. In this case, Figure 2.b shows how solar generation completely displaces the use of diesel during daylight hours, while the diesel generator is mainly used to meet the remaining demand during night hours. The reduced storage capacity follows the same charge-discharge patterns as in the fully renewable system, and the considerably lower PV capacity installed results in reduced overgeneration during the day. The use of diesel remains integral in this system, with diesel fuel costs accounting for 73% of total lifetime costs and 92% of total lifetime GHG emissions.

We observe that diesel usage during day hours can be conveniently and cost-efficiently displaced by solar generation alone. However, large capacities of solar generation and battery storage are needed to completely displace diesel usage during the night while providing high supply reliability. This makes the option of using diesel over night more economical over the first 5-year assessment period evaluated in CLOVER and highlights an impact of short-term budget cycles typical of humanitarian agencies. Larger renewable systems also incur higher levels of overgeneration during the day and higher levels of dumped electricity. This excess generation could potentially be made available to power additional loads and activities during the day, such as productive uses and businesses.

## 3.3. Economic and environmental benefits of sustainable mini-grid systems

Technical, economic and environmental dimensions need to be considered in the decision-making process to select the most suitable system design. Figure 3 shows the net savings over the 15-year system lifetime in total costs, GHG emissions and diesel fuel used, for the hybrid and fully renewable systems presented in Section 3.2 in comparison with the incumbent diesel system in Nyabiheke, considering a reliability level of 95% in all cases. A detailed breakdown of the main economic and environmental metrics of each system is presented in Table 2.

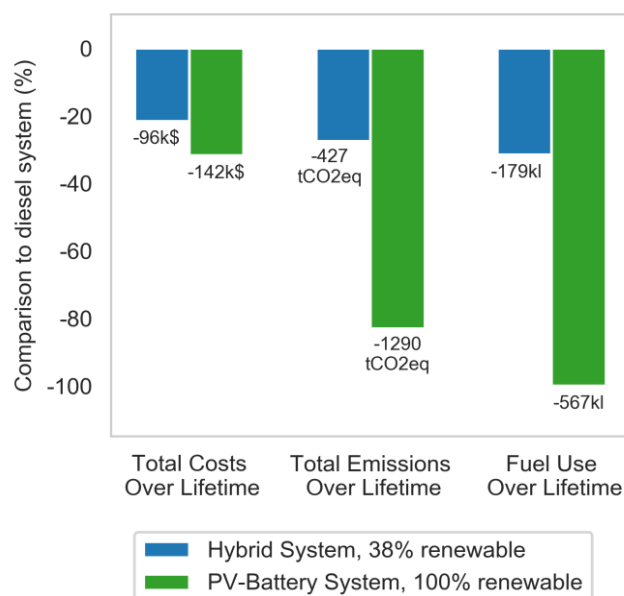

*Figure 3. (Single column color artwork)* Total costs, total GHG emissions and total diesel fuel use savings over system's 15-year lifetime for the optimised PV-battery and hybrid mini-grid designs compared to the incumbent diesel system present in Nyabiheke.

|  | Incumbent diesel system | Optimised hybrid system | Optimised PV-Battery system |
|---|---|---|---|
| Installed PV capacity (kWp) | - | 30 | 100 |
| Installed storage capacity (kWh) | - | 0 | 410 |
| Diesel fuel usage (l) | 567,000 | 388,000 | 0 |
| Total lifetime cost (Thousands of $) | 446 | 350 | 305 |
| LCUE ($/kWh) | 0.557 | 0.409 | 0.353 |
| Total lifetime emissions (tCO2eq) | 1554 | 1127 | 264 |
| Emissions intensity (gCO2/kWh) | 1097 | 719 | 167 |
| Additional initial equipment cost (Thousands of $) | - | 13 | 227 |
| O&M cost without fuel (Thousands of $/year) | 5 | 6 | 6 |
| Diesel fuel cost (Thousands of $/year) | 47 | 31 | 0 |
| Payback of additional new equipment cost | - | 0.9 years | 6.2 years |

*Table 2.* Characteristics of optimised PV-battery and hybrid systems compared to incumbent diesel system for the existing energy demand in Nyabiheke, considering a system lifetime of 15 years with a reliability level of 95%.

Figure 3 shows how both sustainable mini-grid designs present lower lifetime costs and emissions than the diesel system. This is due to the partial or full displacement of diesel fuel use, which involves high operating costs and environmental impact. The fully renewable system yields the best economic and environmental performance in the long term, presenting the lowest LCUE, total lifetime costs (reduced by 32%) and total lifetime emissions (reduced by 83%), while completely displacing the use of diesel fuel compared to the incumbent system. However, to unlock these long-term benefits, an additional initial investment of $227,000 to deploy the system is needed, which is paid back in 6.2 years due to offsetting diesel fuel costs. This upfront cost represents a significant barrier for the deployment of this type of system, considering the funding shortage and short budgeting cycles under which humanitarian originations frequently operate.

The hybrid system presents diminished but still significant savings in total lifetime costs (22%) and fuel use (33%) compared to the incumbent diesel system. Despite this, the displacement of diesel use during daylight hours without relevant storage requirements only requires an additional initial investment in new equipment of $13,000, which presents a payback time of 0.9 years. This investment may be more accessible to the overstretched financial resources available for humanitarian organisations. When considering the cumulative total cost of both sustainable designs, the hybrid system remains the most economical solution during the first 9.8 years of operation, after which it is overtaken by the renewable system.

As introduced in Section 2.2, a minimum reliability level of 95% over the system lifetime is considered for this study. Furthermore, a sensitivity analysis of the system reliability requirements was carried out and can is available in the Supplementary Information. This analysis shows that the reliability requirements have a reduced impact in the LCUE of the diesel and hybrid systems as a result of the generation flexibility provided by diesel generators. However, for the fully renewable system, the additional generation and storage capacity required to meet the demand with increasing reliability requirements translates into a significant increase of the final LCUE of the system. This increase is steady for reliability levels under 95%. Nevertheless, for higher reliability levels, the variability in the solar resource requires progressively more significant increases in the system capacity, resulting in pronounced increases in upfront costs and the LCUE of the system. For reliability levels over 99%, the LCUE of the fully renewable system exceeds the LCUE of the corresponding hybrid system. Therefore, we consider a reliability level of 95% as the best compromise between the security of supply and the cost implications for the system design process.

### 3.4. Sensitivity analysis to the renewable fraction

The previous analysis features a hybrid system, with a renewable fraction of 0.38, which yields the lowest LCUE over the first assessment period of 5 years. This is accompanied by a fully renewable system (renewable fraction of 1), which yields the lowest LCUE over the 15-year system's lifetime. This event highlights the timescale dependence of the system net costs, and the importance of the timeframes considered in the system design choices. These two systems might serve as a simplistic representation of short and long-term approaches that humanitarian actors can take to the design of sustainable mini-grid systems. Nonetheless, there is a continuous spectrum of hybrid systems with renewable fractions between 0.38 and 1, whose cost structures and performance could better suit the financial resources and environmental objectives of specific organisations operating in displacement settings.

Under the same optimisation conditions, Figure 4 illustrates the economic and environmental performance of a range of hybrid systems optimised for different renewable fractions from 0.38 to 1, considering a 95% reliability level over a 15-year lifetime for all. The complete spectrum of hybrid systems and the fully renewable system present negative MACs. This indicates that they can all meet the electricity demand requirements with both lower costs and emissions than the incumbent diesel system over their lifetime. Lower renewable fraction hybrid systems present more negative MACs, meaning that they bring greater cost reductions per tonne of emissions offset. The MAC of hybrid systems experiences a diminishing return at higher renewable fractions, due to the cost associated with the increasing storage capacity needed to displace higher fractions of diesel used overnight. Therefore, lower renewable fraction hybrid systems appear as more cost-effective offsetting diesel use emissions than more renewables-dominated systems.

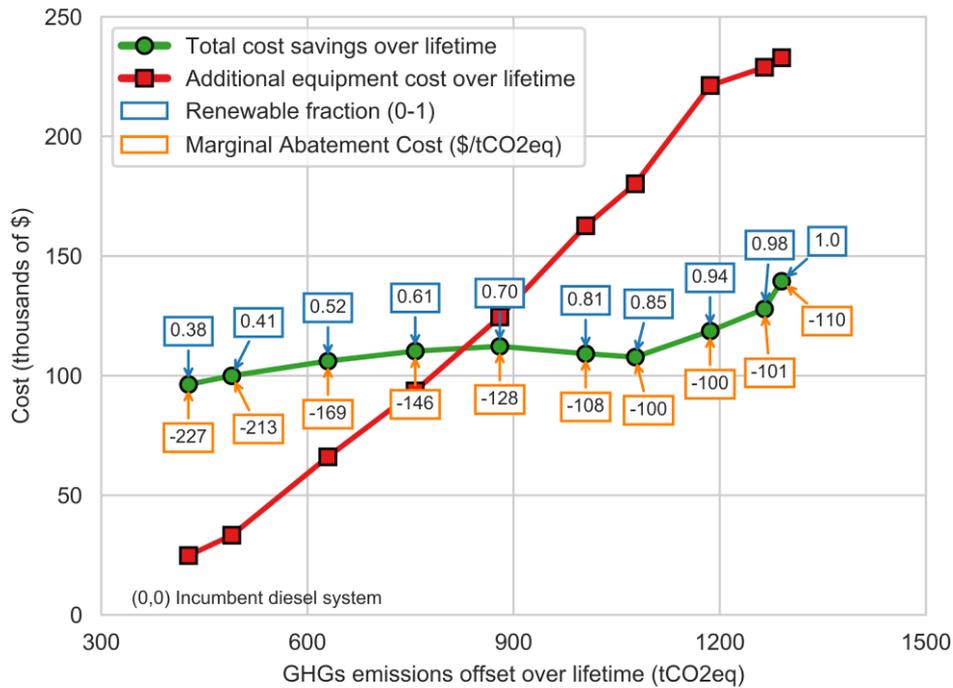

*Figure 4. (1.5 column color artwork)  Total cost savings over lifetime (green line), additional new equipment cost (red line), GHGs emissions offset over lifetime (horizontal axis) and marginal abatement cost (orange box) for different renewable fractions (blue box) of hybrid systems optimised for the existing load in Nyabiheke.*

Figure 4 also represents the trade-off between long-term objectives and short-term constraints characteristic of humanitarian assistance. The long-term objectives include the provision of clean and affordable electricity supply [13], represented by the highest lifetime GHGs emissions savings and total lifetime cost savings possible. The short-term constraints are associated with the difficult access to capital by most humanitarian organisations and represented by the high upfront investments required to introduce capital-intensive renewable energy assets. It is worth highlighting that any level of hybridisation of the system, accounting for the additional investment required in renewable assets, will bring total cost savings over the lifetime of the system in comparison with the incumbent diesel system. This is the case even when the additional upfront costs are higher than the final lifetime savings, being the former accounted for in the latter.

This trade-off between short-term investment barriers and long-term savings provides a useful comparison to guide decision making according to the resources and objectives of humanitarian actors. The system design process in displacement contexts is heavily influenced by the financial resources available, the project timeframe and the capacity of involved actors to handle the asset management risks. It is widely acknowledged that fully renewable systems would bring the highest cost and emissions savings in the long term for humanitarian actors, but have associated much higher upfront costs. Depending on access to capital available for organisations, hybrid systems can bring significant mid and long-term emissions savings in a more cost-effective manner than fully renewable systems, also benefitting of reduced initial investments required.

For the existing electricity demand in Nyabiheke, the three following systems can be identified as potential solutions considering different levels of access to capital for humanitarian organisations involved:

- *Very low access to capital*: Minimum LCUE hybrid system over 5 years with renewable fraction of 0.38, representing an additional initial investment of $13,000 compared to the incumbent diesel system, with a payback of 0.9 years.
- *Moderate access to capital*: Higher renewable fraction hybrid system, 0.7, representing an additional initial investment of $106,000 with a payback of 4.2 years.

- *High access to capital*: Fully renewable system, where an additional initial investment of $227,000 yields a maximum of $142,000 of total savings during the system lifetime, with a payback of 6.2 years.

### 3.5. Operational implications of the technology mix

Despite of the central role of the cost implications of the technology mix and level of hybridisation, the associated technical implications can have a significant impact on the operational sustainability of the projects. The complex nature of humanitarian and displacement settings also presents multiple operational constraints and challenges that influence the system design further than its purely economic and environmental performance. For instance, for hybrid systems, the hours of operation of the diesel generator might be constrained for technical or noise pollution reasons. Such a situation would occur when the system is not capable of running reliably for more than a specified period per day, or when humanitarian organisations try to reduce the disruption caused by noisy diesel generators during specific hours, such as at night.

We evaluate the probability of using the diesel generator for each hour of the day for different hybridisation levels. Higher renewable fractions directly translate into fewer hours of diesel generator operation per day. For instance, a 0.38 renewable fraction system utilises the diesel generator an average of 16.6 hours per day, while these are reduced to 7.8 hours for a 0.7 renewable fraction system and 1.8 hours for a 0.95 renewable fraction system. It can be highlighted that while more diesel-based systems switch on diesel generators earlier in the evening, most systems operate the diesel generator until similar hours in the morning, corresponding to sunrise. Considering that diesel generators are frequently oversized, the renewable fraction can also impact the load factors at which they operate. This can result into poorer fuel efficiencies and higher fuel consumption levels. More information about the impact of different renewable fractions in the asset utilisation is available in the Supplementary Information.

Other events such the seasonal variability of the renewable resource can also affect the output of the renewable system, as well as its degradation over time. These can therefore impact the diesel supply requirements in the case of hybrid systems, with the corresponding impact on the logistical and operational activities of humanitarian actors. In the case of fully renewable systems where no diesel generation is available, these events would translate into shortages in the reliability level if not adequately accounted for in the system design phase. Most practitioners are aware of the technological degradation of PV and battery storage systems over time, and these processes are accounted for in the optimisation process used in CLOVER. The re-assessment of the system and the capacity replacement required to maintain the desired system performance over its lifetime are included in the previous analysis. However, there is little understanding about the operational impact of the seasonal variability of renewable resources in the reliability of renewable mini-grid systems, or in the case of hybrid systems, about the impact on the diesel supply requirements to reliably meet the demand.

In Nyabiheke, situated in Northern Rwanda, the solar energy resource experiences a variation of approximately +/-15% during the main dry season running from April until October and a smaller dry season occurring in December. During the rainy seasons, from October to December and January to April, the solar energy supplied can be 10% lower than average. Over the year, the variation in solar resource is counterbalanced by the reverse variation in the diesel energy supplied by the system to maintain the reliability of the electricity supply. The corresponding diesel fuel supply variations are dependent on the renewable fraction of the system: this ranges from +/-5% for low renewable fraction systems to +/-20% for very high renewable fractions. However, the monthly variation of diesel supply in litres is relatively small across all hybridisation levels, enclosed in a range of +/-80 litres range per month. More information about the impact on diesel supply requirements is available in the Supplementary Information. Although the renewable fraction is not a critical factor in the variability of the amount of diesel being supplied, this issue must be considered in the general operations of humanitarian organisations to ensure the reliability of the electricity supply: if not, this could affect diesel supply contractual agreements, for example.

### 3.6. Mini-grid design approaches for the expansion of energy provision

#### 3.6.1. Short-term design approaches in humanitarian contexts

Considering the current efforts to expand the electricity supply in refugee camps to displaced populations, with a focus on productive and income-generating activities, we introduced three additional demand scenarios in Section 3.1 (Figure 1a). However, despite the short timeframes to which humanitarian organisations are usually bound, we have also highlighted the importance of considering the long-term evolution of the electricity demand in the initial design phases of systems. Short-term planned systems may suffer from inadequate maintenance or unexpected demand growth. Table 3 summarises the main characteristics of the short-term approach, as well as two alternative strategies discussed.

|  | Short-term approach | Long-term static approach | Long-term modular approach |
|---|---|---|---|
| Renewable system size | Undersized, initial deployment | Initially oversized, initial deployment | Adaptative, phased deployment |
| Average diesel supply | Increasing | Increasing, initially lower | Relatively constant |
| Renewable fraction | Decreasing | Initially higher, reducing | High and relatively constant |
| LCUE | Highest, increasing diesel costs | High, large upfront investments | Lowest |
| Wasted energy | Very low | Initially large | Relatively constant |
| Asset utilisation | Very high | Initially low | High |
| Lifetime GHG emissions | Very high | Low | Lowest |

*Table 3. Characteristics short-term (5-year timeframe) and static and modular long-term (15-year timeframe) design approaches for a hybrid system when simulated for a growing load profile over 15 years.*

We simulate the impact of connecting an unplanned growing load over 15 years to different renewable fraction systems, initially optimised to supply a constant load – corresponding to the present load in Nyabiheke – over five years, without any re-assessment or re-sizing of the system after its initial deployment (the "short-term" approach). The final load connected corresponds to High Productive scenario load, equivalent to the additional connection of 166 Tier 3 businesses. The connection of additional productive users considered is progressive over time, proliferating after Year 2 and being completed by Year 10 of simulation.

Under these conditions, the performance over 15 years of a 0.7 renewable fraction hybrid system and a fully renewable system is shown in Figures 5.a and 5.b respectively, labelled as the short-term approach. Two additional long-term design approaches, discussed later, are also included. A reliability level of 95% is considered for all the systems simulated. Figures 5.a and 5.b show the growing load over time, together with the average wasted energy and the level of diesel energy supplied in the different design approaches. For the short-term approach, we observe that the unexpected demand growth over time, in addition to the degradation of the system capacity – initially designed for just 5 years – translates into increases in the diesel energy supply needed to meet the demand maintaining the same level of reliability. This growth of diesel energy supplied exactly matches the demand growth plus the progressive degradation of the system after Year 5 of simulation. As a result, the eventualities not considered in the short-term approach have an impact on the performance of the system compared to its initial functionality.

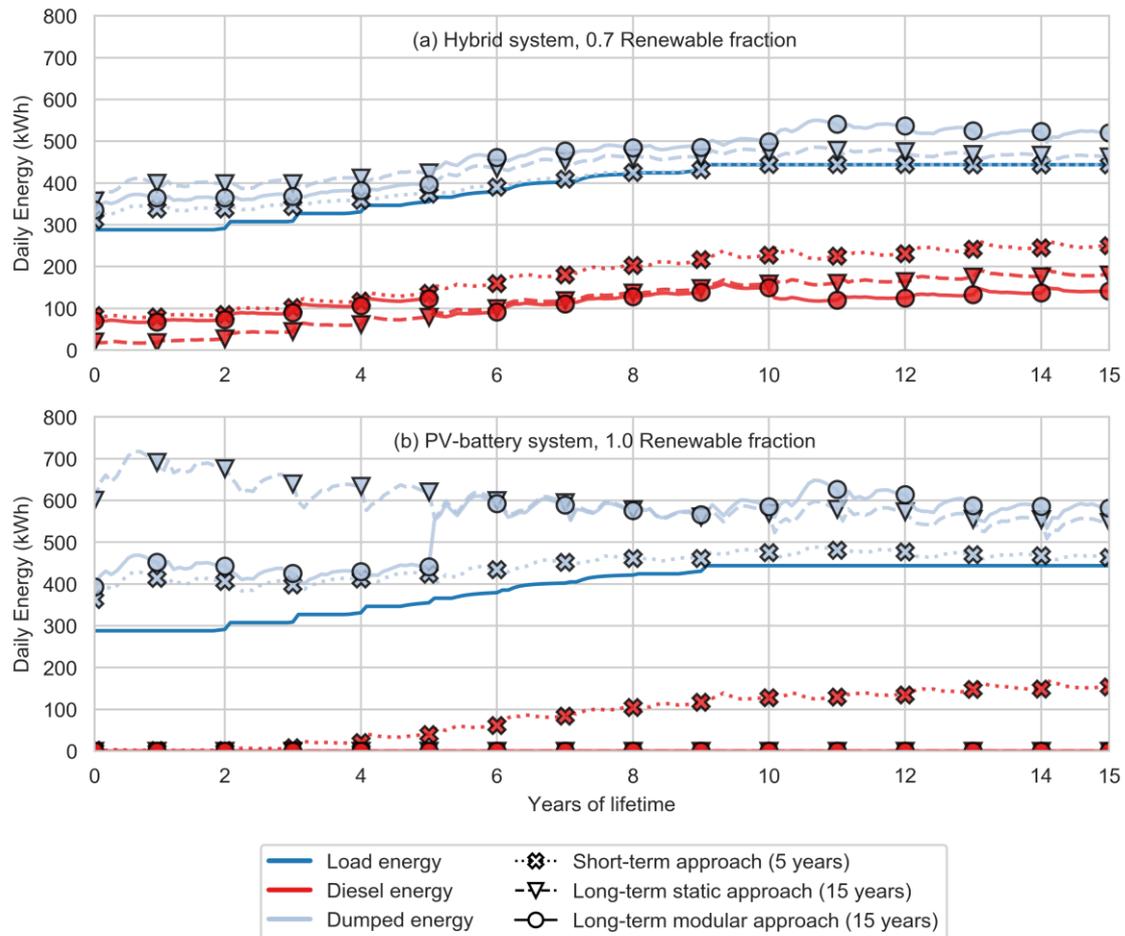

***Figure 5.*** *(Double column color artwork) Average daily demand load, diesel energy supplied and wasted energy for a) a hybrid system with a renewable fraction of 0.7 and b) a fully renewable system under three different design approaches (short-term, static long-term and modular long-term) when meeting a growing load over 15 years.*

For instance, under the short-term approach a system designed to have a renewable fraction of 0.7 for the first five years falls to just 0.54 over its entire lifetime. This increases the LCUE of the system by 8% and the emissions intensity of the electricity supplied by 37% as a result of the higher diesel usage to meet this unexpected load growth. Due to the operational flexibility of diesel generators, the system can absorb the increase in demand over time if sufficient diesel generation capacity is in place: this case is common in displacement settings, where oversized diesel generators have been the norm to date. Unexpected load growth does not directly translate into a loss of reliability in the electricity supply, but significantly worsens the economic and environmental performance of the system. Additionally, logistic and operational challenges associated with varying levels of diesel supply needed might arise over its lifetime.

In the case of the fully renewable system, represented in Figure 5.b, the unexpected connection of additional loads would translate into a drop in reliability. However, humanitarian actors in charge of energy provision in camps would likely not allow the connection of these loads if the security of the supply were at risk. In a context of lack of access to capital to expand the renewable system, the most straightforward solution is the progressive (re-)introduction of diesel generation over time to meet the demand growth while maintaining the system reliability.

Figure 5.b illustrates this event under the short-term approach. After 15 years, the resulting increase of LCUE and emissions intensity compared to the initially planned system for a five-year timeframe are of 17% and 120% respectively, reducing the final renewable fraction from 1.0 to 0.78. This can pose challenges in cases where the diesel generator is removed or not present

due to the initial aim of completely supplying the demand through clean energy sources. It also highlights the value of the flexibility that traditional diesel generators can offer when complementing cleaner energy technologies to ensure more resilient electricity supply.

### 3.6.2. Options for long-term design approaches

These observations call for the design of sustainable mini-grid systems in displacement settings to be conducted accounting for eventualities such as the increase of the electricity demand over time. Despite the short-term nature of humanitarian assistance, this reinforces the need to consider longer timeframes in system design phases to ensure that the system performance falls within operational and financial capabilities of organisations in the long-term.

To achieve this, we considered two different approaches to design a system that meets a growing demand with a required level of reliability over the entire lifetime of the system:

- *Static approach*: A system deployed in a one-off installation and designed to cover the final demand requirements at the end of the system lifetime.
- *Modular approach*: A modular system, re-assessed and re-sized at specific points to adapt to the growing demand, introducing the additional generation and storage capacity required to meet the existing load during each period.

As mentioned earlier, Figure 5.a also displays both long-term approaches for a hybrid system optimised to meet the growing demand profile from the Baseline scenario to High Productive scenario with a reliability level of 95% and a renewable fraction of 0.7. It shows the average wasted energy and the level of diesel energy supplied daily for both approaches. The corresponding approaches for a fully renewable system with a renewable fraction of 1.0 are shown in Figure 5.b.

For the fully renewable system, the static approach results in much higher wasted energy levels (+25%) in comparison with the modular approach. This increase is due to the need to have an initially hugely oversized system to account for the system degradation over time in addition to the demand growth. The oversizing of the static system results in inferior technical performance and asset utilisation, leading to higher costs (+25% total lifetime costs) compared to the modular system. A complete table with the system performance comparison for both long-term approaches is available in Tables A.3 and A.4 in the Appendix.

For the 0.7 renewable fraction hybrid system, the static approach presents 5% higher cumulative costs and 12% higher new equipment costs over 15 years. However, despite having the same levels of diesel fuel used, the average levels of wasted energy are 14% lower than in the modular approach. This is because, although the static approach presents an initially oversized system, the level of overgeneration is significantly reduced as the load grows and the system degrades over time, compared to the modular counterpart. These differences reinforce the importance of undertaking comprehensive energy demand assessments to evaluate growing electricity demand and the potential connection of productive users. They also advocate for a multi-stage approach that allows the expansion of the system alongside the demand growth.

The sizes of the systems discussed, together with the analogous analysis presented here for a growing demand profile from the Baseline scenario to the High Productive scenario, are available for the Low Productive and Medium Productive scenarios in the Supplementary Information.

### 3.6.3. Socio-economic implications of connecting productive users

The socio-economic benefits of expanding the electricity supply to productive users in camps are widely promoted. However, the viability of the additional investments required is crucial in the current climate of funding shortages and perceived risk for the engagement of the private sector in the energy supply in displacement settings. The need of recovering such additional investments collides with the aim of providing electricity to refugee prices at an affordable price, without giving them a favour in comparison to surrounding host communities, which can generate undesirable tensions. If the renewable energy infrastructure to power humanitarian operations is introduced, the expansion required to extend the electricity supply to refugees does not necessarily compromise the cost-effectiveness of the system nor the profitability for private sector players, assuming that the system is suitably designed and implemented.

The load profile of productive users, who concentrate their activity during day hours (Figure 1b), generally matches well with the solar generation profile. This pairing proportionally reduces the need for additional storage and can improve the overall cost-effectiveness of the system. We have calculated the payback times of the systems presented in Figure 5 to power additional businesses compared with the use of diesel generators only to power the growing demand. Our analysis shows that paybacks of modular, sustainable systems designed for progressively connecting productive users – resulting in a more day-centred load profile – are shorter than the corresponding paybacks of systems designed only to meet a constant demand profile, like the institutional demand presented in Section 3.1. These payback times are reduced from 4.2 years to 3.3 years for the case of the 0.7 renewable fraction system, and from 6.2 years to 6.0 years for the fully renewable system. For the static approach, however, the initial oversizing of the system and the lower asset utilisation translate into significantly longer paybacks, of 6.6 and 10.2 years for the hybrid and fully renewable systems respectively.

In order to prevent tensions with surrounding host communities while ensuring the financial sustainability of the projects, refugee business should pay a fair price for their electricity supply. A reasonable case is to consider that the connected productive users are charged at the current Rwanda national grid tariff (204 RWF/kWh or 0.224 $/kWh). Under this assumption, a significant fraction (69%) of the additional costs of the modular system designed to connect the productive users compared to the initial hybrid system designed only for the existing institutional loads in Nyabiheke can be recovered in 15 years, with a shortfall of only around $29k. On the other hand, for the static approach system, the higher upfront investments required results in a lower return of the additional investment associated over the 15 years, with only 57% recovered and a shortfall of $50k. This difference is even starker for the fully renewable system, where the discounted revenue from productive users connected over 15 years increases to 85% (with a $11k shortfall) and 37% (with a $109k shortfall) of the additional investment required to deploy the expanded modular and static systems respectively.

Humanitarian actors should consider the compatibility of the desired system characteristics and design approaches with the existing funding structure and the resources available. While the modular approach might suit better long-term sustainable practices, it requires closer monitoring and phased deployment. The static approach can adjust better to contexts where one-off funding is provided by donors at the initial stage of the project, and consequent interventions are less likely. The search for pathways to unlock longer-term and more cost-effective solutions are set to play a crucial role in the transition of humanitarian assistance towards a cleaner energy supply.

### 3.7. Mini-grid delivery models in humanitarian settings

As presented in Section 2, the engagement and establishment of contractual mechanisms with the private sector is crucial in the successful scale-up of sustainable energy solutions in humanitarian settings. This section presents a preliminary cash flow analysis of the cost implications for humanitarian organisations associated with the deployment of the sustainable mini-grid systems presented in Section 3.4, under the previous delivery models outlined. Although the provision of clean electricity to promote productive activities is desirable and potentially profitable, as shown in Section 3.6, this analysis focuses on the introduction of renewable technologies to power humanitarian operations only as a first step towards the provision of broader clean energy supply in humanitarian settings.

The system selected for this analysis is the hybrid system with a renewable fraction of 0.7. Table 4 presents the discounted cashflows assumed by humanitarian organisations for different delivery models compared to the operation of the current diesel system. These models include purchase, lease to own and limited PPA. As previously mentioned, humanitarian organisations have a limited capability to undertake long-term investments. Therefore, different timeframes for the lease to own and PPA agreements are studied to highlight the potential benefits of committing to longer-term agreements.

|                    | Baseline Diesel System | Purchase | Lease to own (5 years) | Lease to own (10 years) | Limited PPA (5 years) | Limited PPA (15 years) |
|---|---|---|---|---|---|---|
| CAPEX Year 1       | 0   | 107 | 0   | 0   | 0   | 0   |
| CAPEX Year 5       | 0   | 13  | 13  | 0   | 0   | 0   |
| CAPEX Year 10      | 0   | 5   | 5   | 5   | 0   | 0   |
| Annual OPEX        | 52  | 23  | 57  | 55  | 45  | 44  |
| Total Cost Year 5  | 214 | 198 | 238 | 227 | 188 | 183 |
| Total Cost Year 10 | 344 | 268 | 308 | 367 | 303 | 295 |
| Total Cost Year 15 | 423 | 310 | 350 | 409 | 375 | 364 |

*Table 4. Discounted cashflows assumed by humanitarian organisations for different delivery models proposed to implement a hybrid system optimised for the existing energy demand in Nyabiheke with a renewable fraction of 0.7, compared with the cashflows associated to the use of the incumbent diesel system over a period of 15 years.*

Discounted cashflows assumed by organisations calculated considering different rates of return on investment (ROI) for the energy services company under the current Rwandan Corporate Tax Rate of 30% [53] and an interest rate of 9.5% [34,54]. For the lease to own of five years, we consider a 20% ROI. A 40% ROI is used for the lease to own of 10 years. The limited PPA of five years is designed to recover 95% of the costs of the system deployed by Year 5 for the energy services company. This agreement can be recurrently extended over the 15-year lifetime of the system at the same rates. The Limited PPA of 15 years considers an ROI of 20% for the energy services company.

For the 0.7 renewable fraction system, all of the proposed delivery models present lower total lifetime costs over 15 years than the incumbent diesel system. Besides, they also provide the environmental benefits of displacing 70% of diesel use, both in terms of GHG emissions and local pollution levels. The purchase option presents the lowest total lifetime costs, but the highest upfront investments required.

In contrast, lease to own and PPA models remove the need for initial investments on renewable assets in exchange of a monthly fee. Monthly fees and annual costs under lease to own and PPA agreements are significantly higher than the ones corresponding to the purchase of the system. In return, the operation and maintenance of the renewable assets are fully outsourced to the energy services company, removing the operational risk from humanitarian organisations.

Furthermore, for PPA models, the annual operational costs are lower than the costs of operating the incumbent diesel system. For lease to own models, despite being slightly higher, they present the advantage of acquiring the ownership of the renewable assets after the end of the lease.

If organisations are interested in the ownership of the system in the long-term, the lease to own of five years presents 9.5% higher annual costs than the operation of the diesel system during the first five years. The lease to own of 10 years reduces this increase to 5.7%, further benefiting from five additional years of outsourced renewable energy provision. However, the ownership and operation of large and unmovable assets is often challenging for humanitarian organisations, whose risks are intensified by the temporary nature and the long-term uncertainty perceived around displacement settings.

If organisations are not interested in the ownership of the system or the in-house management of the energy supply, PPA models present an economical and practical solution. A recurrent five-year Limited PPA presents 13% lower costs than the operation of the incumbent diesel system and a 15-year Limited PPA brings further annual cost reductions up to 15%. Alongside cost savings, organisations benefit from fully outsourced renewable energy provision without the risk associated with asset ownership and management in the long-term.

Table 4 shows how the possibility of committing to longer-term agreements, both through lease to own or PPA models, can reduce the annual costs of introducing sustainable mini-grid systems. Moreover, these longer-term agreements would facilitate the engagement of private energy services providers by providing greater certainties about the returns on their investments.

The corresponding analysis for the hybrid system with a renewable fraction of 0.38 and the fully renewable system is available in Tables A.5 and A.6 in the Appendix. For lower renewable fractions, where the size of the renewable energy assets installed is more reduced, shorter-term agreements such as a lease to own of five years or recurrent five-year PPA agreements facilitate the introduction of renewable generation at the lowest cost possible, close to the costs of purchasing and operating the system by organisations. For higher renewable fraction systems, the longer-term agreements allow the outsourcing of the higher-risk system operation and energy supply. PPAs appear as an economical and efficient solution to remove the risk associated with the ownership and management of large renewable assets in displacement settings.

It is worth noticing that the results of this analysis are susceptible to specific parameters such as the cost of the diesel fuel. This study considers the national petrol price for Rwanda (1.192 $/l) (36). However, the remote location of many refugee camps often involves significantly high costs to transport and supply the diesel to these isolated locations. Table A.7 in the Appendix represents the cashflow implications of a 0.7 renewable system considering a 20% diesel price increase. As expected, the increased fuel cost further improves the business case for the introduction of renewable systems in humanitarian contexts. Furthermore, the benefits that contractual mechanisms such as PPA and Lease to own models bring in comparison to the system purchase are also enhanced under higher diesel price scenarios.

Most successful projects to date have received financial support to overcome the high upfront cost of renewable assets and ensure the profitability of the mini-grids. This support is also key to facilitate the engagement of private sector players. To illustrate this, Table A.8 in the Appendix shows the impact of subsidising 50% of the upfront investments required in renewable assets for a 0.7 renewable fraction system. The financial support can dramatically reduce the costs assumed for organisations while also increasing the profit margins for energy services companies. Public financing is already a valuable tool to introduce and scale-up clean energy solutions in rural off-grid contexts, and it can also help to remove the uncertainty associated with the investment in renewable infrastructure in humanitarian contexts.

## 4. Conclusions and Recommendations

### 4.1. Potential of sustainable mini-grids in humanitarian settings

Despite increasing environmental awareness and recent initiatives from the humanitarian sector, the use of fossil fuels for energy supply in humanitarian operations remains central. The successful deployment of renewable energy solutions has been hampered by the overstretched resources of humanitarian actors and the uncertain regulatory frameworks around humanitarian settings.

We analysed several possible sustainable mini-grid designs for the case study of Nyabiheke refugee camp in Rwanda. When operating under very limited timeframes and financial capabilities, a diesel system might be the most economical and straightforward solution. However, we have shown that fully renewable mini-grid systems, completely displacing diesel use through solar generation and battery storage, provide the best economic and environmental performance in the long-term. For the case of Nyabiheke, these savings could represent up to 32% of total costs and 83% of total GHG emissions over 15 years. Per contra, they involve substantially higher upfront investments and longer payback times, 6.2 years for the case of Nyabiheke. Thus, these paybacks may be beyond the financial capabilities of humanitarian actors who are generally unable to commit to multi-year arrangements and working on tight annual budgets.

As an alternative, the hybridisation of renewable systems using already present diesel generation can provide a successful compromise solution to meet the current electricity needs in displacement settings in more cost-effectively. Hybrid solar-battery-diesel systems benefit from lower upfront investments in renewable assets required, translating in shorter paybacks (some as low as 0.9 years for the case of Nyabiheke) that can better match the timeframes and access to capital of humanitarian actors.

Our analysis shows that (albeit only partially) seizing the environmental benefits of renewable energy, hybrid systems are a more cost-effective way of offsetting diesel GHG emissions when compared to fully renewable systems. This is owed to the lower energy storage requirements that translate into comparatively lower upfront investments, also leading to lower marginal abatement costs for diesel emissions. These were found to be negative over the system lifetime for any fraction of renewable energy introduced in the system, saving both costs and emissions compared to exclusively using diesel as at present. When planning the introduction of hybrid mini-grid systems, the evaluation of multiple levels of hybridisation can also help to adapt to the existing operational constraints in diesel generator operation and fuel supply logistics. Furthermore, it can inform potential operational disruptions in the future.

The introduction of renewable infrastructure in displacement settings also presents an opportunity to expand the electricity supply to displaced populations, that at present is widely inexistent or inadequate. Aligned with current efforts to integrate displaced communities in the socio-economic life of host countries, we provide evidence on how the benefits of the expansion of the electricity supply to productive users. This expansion can improve the asset utilisation and cost-efficiency of the overall system in addition to generating economic and livelihoods opportunities for displaced communities. This can result in shorter payback times of the initial investment (from 4.2 to 3.3 years for a 0.7 renewable fraction system and from 6.2 to 6.0 years for a fully renewable system). This is due to the favourable match of the productive activities load profiles with the solar generation profile.

During design phases it is crucial to account for the potential expansion of the load connected to the system over its lifetime. This prevision is essential to ensure appropriate performance and the security of the supply in the long-term. To maximise the benefits of the extension of the electricity supply, undertaking detailed and recurrent energy assessments is fundamental. Continuous assessments allow the adaptation of the system capacity to the potentially growing electricity demand, avoiding widely oversized systems and ensuring a stable performance over the project lifetime. For the case of Nyabiheke, a one-off initial installation of a fully renewable system can present 20% higher lifetime costs and 25% lower asset utilisation when compared to a five-year recurrent system expansion.

In cases where the existing regulatory framework can provide certainty around the long-term status of displacement settings, and where humanitarian organisations can commit to multi-year arrangements, the engagement of private sector actors can play a crucial role in the deployment and scale-up of these solutions. This can be achieved through different contractual mechanisms that can mutually benefit humanitarian players and energy services providers. The outsourcing of the electricity supply removes the upfront financial barriers and reduces the asset management risk from humanitarian players, benefitting from the technical and financial capacity of private actors. These, in turn, receive a stable revenue source and access to a potentially large refugee customer base, maximising the financial and social sustainability of the projects.

### 4.2. Recommendations for practitioners, private actors and policy-makers

The analysis presented in this work can help humanitarian agencies, private actors and policy-makers to quantify the widely acknowledged benefits that renewable technologies can bring in humanitarian contexts, contributing to a growing evidence base about the potential of these solutions. Additionally, it can also throw light on possible strategies to address the existing barriers that have prevented a successful scale-up of these solutions.

Beyond the close cooperation of the aforementioned stakeholders, an underpinning condition to unlock a widespread adoption of cleaner energy in displacement settings is the establishment of favourable policy and regulatory frameworks. These are key to de-risk the investment in long-term energy infrastructure in protracted humanitarian contexts. International agreements such as the CRRF and improved refugee rights provide a base for the integration of displaced communities in the socio-economic life of host countries. However, their explicit introduction in national electrification agendas remains crucial to provide them with equitable livelihood opportunities. Both in displacement and rural electrification terms, Rwanda presents a noticeably progressive framework, and some efforts could be replicated in other host countries.

For humanitarian actors who want to introduce renewable technologies to power their operations, the most suitable system design will depend on their specific financial resources and environmental objectives. In cases where there is no previous energy infrastructure and the logistics and cost of diesel supply are challenging, fully renewable systems present the cleanest and most economical solution in the long-term, given the access to the required capital. This is also the case where the continuity of the project in the foreseeable future is expected. On the other hand, detailed energy demand assessments and the access to the required technical expertise will be even more crucial to ensure that reliability of the electricity supply is not affected over time due to system failures or load growth.

Hybrid systems appear as a more appropriate option where diesel generators are already present, and organisations have limited capacity to face up-front investments. Their shorter payback periods also make them more convenient to interventions with a limited timeframe. In this case, the most appropriate level of hybridisation - whether it involves the displacement of daytime or higher fractions of diesel usage - needs to be carefully assessed. Considerations as the existing financial resources, environmental objectives and the operational implications on the diesel supply and use will determine the optimal technology mix.

We have shown that humanitarian actors can benefit from the engagement of the private sector through the temporary or permanent outsourcing of their energy supply, taking advantage of the private sector technical expertise needed to successfully design, operate and manage renewable energy assets. Tailored contractual mechanisms can provide a best-fit solution for the varying interests and resources of organisations. Lease to own models are suitable for organisations who have a long-term or permanent presence in these settings and want to acquire ownership of the assets. They are particularly cost-effective for reduced renewable capacities installed. PPA models are more advantageous for capital-intensive higher renewable fraction systems, reducing the risk associated with the ownership of large long-life assets in camps. Longer-term agreements can reduce the monthly fees assumed by humanitarian actors. This is especially remarkable for high renewable fractions, by paying off the high upfront cost of the energy assets over a longer timeframe, whilst providing longer revenue certainty for private actors.

In this sense, a restructuring of funding mechanisms and donor practices is also needed to allow multi-year financing initiatives and facilitate the engagement of private actors. Donors who can provide capital can drastically reduce the cost of these contractual mechanisms for humanitarian organisations by subsidising upfront costs. This can also increase the profit margins for private actors and thus make their participation more appealing and less risky. We have also shown the advantages of allocating donor's capital to phased deployments and recurrent energy demand assessments, instead of one-off deployments. This approach can improve asset utilisation, avoiding initially oversized systems and better adapting to future energy supply requirements.

For private sector energy providers entering humanitarian settings, the search of alternative financing sources can facilitate addressing the upfront investments and mitigating the risk of investing in displacement settings. These include peer-to-peer lending and crowdfunding, in addition to partnerships with humanitarian organisations to access development funds. When willing to expand their energy supply to displaced communities, scoping assessments of energy needs to identify the potential customer base of productive users are crucial to informing appropriate system expansions. Similarly, facilitating financing and access to productive appliances can stimulate local energy demand and improve the profitability of mini-grids.

Energy provision is increasingly being recognised as a critical enabler of humanitarian assistance and humanitarian organisations are pledging to introduce cleaner energy sources in their operations. However, higher levels of coordination and the diffusion of data and evidence about the potential of clean energy technologies in humanitarian settings among private actors and governments is needed. Through this work, we have provided an initial assessment of the substantial benefits of renewable mini-grids in these contexts. We also call for more extensive data collection and comprehensive case-specific studies required to promote and generalise these solutions.

A. Appendix

| Modelling parameters | Value | Units | Notes |
|---|---|---|---|
| Optimisation period | 15 | years | |
| Re-assessment intervals | 5 | years | |
| Sufficiency criteria | 0.05 | - | Blackout percentage |
| Optimisation criteria | - | $/kWh | System LCUE |
| Technologies available | - | - | PV, diesel, battery storage |
| PV initial capacity | Various | kWp | Dependant on system |
| PV capacity step (optimisation) | 5 | kWp | 5 kWp resolution |
| Storage initial capacity | Various | kWh | Dependant on system |
| Storage capacity step (optimisation) | 5 | kWh | 5 kWh resolution |

*Table A.1.* Main modelling inputs used for the analysis in CLOVER.

| Technical parameters | Value | Units | Notes |
|---|---|---|---|
| PV azimuth | 180 | ° | South facing |
| PV tilt angle | 10 | ° | From horizontal |
| PV lifetime | 20 | years | From [43] |
| Battery depth of discharge | 50 | % | From [55] |
| Battery round-trip efficiency | 90 | % | From [55] |
| Battery C rate | 0.33 | - | From [56] |
| Battery lifetime | 1500 | cycles | From [55] |
| Diesel generator consumption | 0.4 | l/h per kW | From [57] |
| Diesel generator minimum load factor | 35 | % | From [57] |
| Power conversion efficiency | 95 | % | From [43] |
| Network transmission efficiency | 95 | % | From [56] |

| Financial parameters | Value | Units | Notes |
|---|---|---|---|
| PV module | 500 | $/kWp | From [47,58] |
| PV module O&M | 15 | $/kWp p.a. | From [56] |
| Battery storage (Lead-acid) | 350 | $/kWh | From [55], higher range |
| Battery storage O&M | 10 | $/kWh p.a. | From [56] |
| Battery Operating System | 200 | $/kW | From [59] |
| Diesel generator | 560 | $/kW | From [60] |
| Diesel fuel | 1.192 | $/l | From [61] |
| Diesel generator O&M | 350 | $/Kw p.a. | From [60] |
| PV annual cost decrease | 3 | % p.a. | From [47,58] |
| Diesel fuel annual cost decrease | 0 | % p.a. | Based on [61] |
| Discount rate | 9.5 | % p.a. | From [62] |

| Environmental parameters | Value | Units | Notes |
|---|---|---|---|
| PV module | 1400 | kgCO2eq/kWp | From [63] |
| Battery storage | 110 | kgCO2eq/kWh | From [64] |
| Battery Operating System | 200 | kgCO2eq/kWh | From [65] |
| Diesel generator | 2000 | kgCO2eq/kW | From [66] |
| Diesel fuel | 2.68 | kgCO2eq/l | From [66] |

*Table A.2.* Technical, financial and environmental inputs used for the analysis in CLOVER.

|                                                   | Modular approach system | Static approach system |
|--------------------------------------------------:|:-----------------------:|:----------------------:|
| Average wasted energy (kWh/day)                   | 99                      | 85 (-14%)              |
| Average PV utilisation (kWh.day/kWp installed)    | 3.16                    | 3.04 (-20%)            |
| Average storage utilisation (kWh.day/kWh installed)| 0.42                   | 0.33 (-21%)            |
| Total lifetime cost (Thousands of $)              | 431                     | 452 (+5%)              |
| Total new equipment cost (Thousands of $)         | 189                     | 212(+12%)              |
| Average annual OPEX (Thousands of $/year)         | 16                      | 16 (-1%)               |
| LCUE ($/kWh)                                      | 0.392                   | 0.409 (+5%)            |
| Payback of additional investment vs diesel system | 3.3                     | 6.6                    |
| Total lifetime GHG emissions (tCO2eq)             | 962                     | 954 (-1%)              |

*Table A.3.* Technical performance and economic metrics comparison of static and modular design approaches for a 0.7 renewable fraction hybrid system optimised to meet a growing demand profile over its lifetime, from the Baseline scenario to the High Productive scenario load profile.

|                                                   | Modular approach system | Static approach system |
|--------------------------------------------------:|:-----------------------:|:----------------------:|
| Average wasted energy (kWh/day)                   | 232                     | 290 (+25%)             |
| Average PV utilisation (kWh.day/kWp installed)    | 2.85                    | 2.28 (-20%)            |
| Average storage utilisation (kWh.day/kWh installed)| 0.39                   | 0.32 (-18%)            |
| Total lifetime cost (Thousands of $)              | 386                     | 484 (+20%)             |
| Total new equipment cost (Thousands of $)         | 324                     | 401 (+24%)             |
| Average annual OPEX (Thousands of $/year)         | 4                       | 5 (+25%)               |
| LCUE ($/kWh)                                      | 0.349                   | 0.436 (+20%)           |
| Payback of additional investment vs diesel system | 6                       | 10.2                   |
| Total lifetime GHG emissions (tCO2eq)             | 396                     | 432 (+9%)              |

*Table A.4.* Technical performance and economic metrics comparison of static and modular design approaches for a fully renewable system optimised to meet a growing demand profile over its lifetime, from the Baseline scenario to the High Productive scenario load profile.

|                    | Baseline Diesel System | Purchase | Lease to own (5 years) | Lease to own (10 years) | Limited PPA (5 years) | Limited PPA (15 years) |
|-------------------:|:----------------------:|:--------:|:----------------------:|:-----------------------:|:---------------------:|:----------------------:|
| CAPEX Year 1       | 0                      | 13       | 0                      | 0                       | 0                     | 0                      |
| CAPEX Year 5       | 0                      | 9        | 9                      | 0                       | 0                     | 0                      |
| CAPEX Year 10      | 0                      | 3        | 3                      | 3                       | 0                     | 0                      |
| Annual OPEX        | 52                     | 37       | 42                     | 45                      | 39                    | 42                     |
| Total Cost Year 5  | 214                    | 164      | 174                    | 186                     | 161                   | 176                    |
| Total Cost Year 10 | 344                    | 265      | 275                    | 298                     | 259                   | 281                    |
| Total Cost Year 15 | 423                    | 326      | 336                    | 359                     | 319                   | 346                    |

*Table A.5.* Discounted cashflows assumed by humanitarian organisations for different delivery models proposed to implement a hybrid system optimised for the existing energy demand in Nyabiheke with a renewable fraction of 0.38, compared with the cashflows associated to the use of the incumbent diesel system over a period of 15 years.

|                     | Baseline Diesel System | Purchase | Lease to own (5 years) | Lease to own (10 years) | Limited PPA (5 years) | Limited PPA (15 years) |
|---------------------|---:|---:|---:|---:|---:|---:|
| CAPEX Year 1        | 0   | 227 | 0   | 0   | 0   | 0   |
| CAPEX Year 5        | 0   | 19  | 19  | 0   | 0   | 0   |
| CAPEX Year 10       | 0   | 9   | 9   | 9   | 0   | 0   |
| Annual OPEX         | 52  | 6   | 71  | 62  | 51  | 44  |
| Total Cost Year 5   | 214 | 252 | 293 | 256 | 211 | 183 |
| Total Cost Year 10  | 344 | 286 | 327 | 412 | 339 | 294 |
| Total Cost Year 15  | 423 | 305 | 345 | 430 | 417 | 360 |

*Table A.6.* Discounted cashflows assumed by humanitarian organisations for different delivery models proposed to implement a fully renewable system optimised for the existing energy demand in Nyabiheke with a renewable fraction of 1, compared with the cashflows associated to the use of the incumbent diesel system over a period of 15 years.

|                     | Baseline Diesel System | Purchase | Lease to own (5 years) | Lease to own (10 years) | Limited PPA (5 years) | Limited PPA (15 years) |
|---------------------|---:|---:|---:|---:|---:|---:|
| CAPEX Year 1        | 0   | 107 | 0   | 0   | 0   | 0   |
| CAPEX Year 5        | 0   | 13  | 13  | 0   | 0   | 0   |
| CAPEX Year 10       | 0   | 5   | 5   | 5   | 0   | 0   |
| Annual OPEX         | 63  | 28  | 60  | 59  | 48  | 54  |
| Total Cost Year 5   | 260 | 210 | 249 | 244 | 200 | 220 |
| Total Cost Year 10  | 418 | 300 | 327 | 393 | 322 | 354 |
| Total Cost Year 15  | 499 | 366 | 374 | 439 | 391 | 429 |

*Table A.7.* Discounted cashflows assumed by humanitarian organisations for different delivery models proposed to implement an hybrid system optimised for the existing energy demand in Nyabiheke with a renewable fraction of 0.7, compared with the cashflows associated to the use of the incumbent diesel system over a period of 15 years and considering a diesel fuel price of 1.43 $/l .

|                     | Baseline Diesel System | Purchase | Lease to own (5 years) | Lease to own (10 years) | Limited PPA (5 years) | Limited PPA (15 years) |
|---------------------|---:|---:|---:|---:|---:|---:|
| CAPEX Year 1        | 0   | 53  | 0   | 0   | 0   | 0   |
| CAPEX Year 5        | 0   | 13  | 13  | 0   | 0   | 0   |
| CAPEX Year 10       | 0   | 5   | 5   | 5   | 0   | 0   |
| Annual OPEX         | 52  | 23  | 41  | 43  | 34  | 36  |
| Total Cost Year 5   | 214 | 145 | 169 | 175 | 139 | 148 |
| Total Cost Year 10  | 344 | 215 | 239 | 282 | 224 | 239 |
| Total Cost Year 15  | 423 | 257 | 281 | 325 | 277 | 296 |

*Table A.8.* Discounted cashflows assumed by humanitarian organisations for different delivery models proposed to implement an hybrid system optimised for the existing energy demand in Nyabiheke with a renewable fraction of 0.7, compared with the cashflows associated to the use of the incumbent diesel system over a period of 15 years, and considering a 50% subsidy of the upfront cost of the hybrid system.

**Acknowledgements**

The authors would like to acknowledge Tracy Tunge, Sarah Begg and Laura Clarke of Practical Action for their contributions to the development of this paper, and the IKEA Foundation for supporting the Renewable Energy for Refugees Project, a partnership between Practical Action and UNHCR. They would also like to acknowledge the support of Laila Read and the Grantham Institute - Climate Change and the Environment.

**Funding Information**

JBA and PS would like to acknowledge Research England GCRF QR Funding, and PS would like to acknowledge the funding received from EPSRC (EP/R511547/1, EP/R030235/1 and EP/P02484X/1).